\documentclass{article}

\usepackage[nonatbib, preprint]{neurips_2022}

\usepackage[utf8]{inputenc} % allow utf-8 input
\usepackage[T1]{fontenc}    % use 8-bit T1 fonts
\usepackage{hyperref}       % hyperlinks
\usepackage{url}            % simple URL typesetting
\usepackage{booktabs}       % professional-quality tables
\usepackage{amsfonts}       % blackboard math symbols
\usepackage{nicefrac}       % compact symbols for 1/2, etc.
\usepackage{microtype}      % microtypography
\usepackage{xcolor}         % colors
\usepackage{float}
\usepackage[backend=bibtex,citestyle=numeric-comp,sorting=none, language=british]{biblatex}
\usepackage{xkeyval}
\usepackage{amssymb,graphicx,amsmath}
\usepackage{caption}
\usepackage{subcaption}

\title{Functional Connectome: Approximating Brain Networks with Artificial Neural Networks}

\author{
  Sihao Liu (Daniel)\\
  Research Department of Cell and Developmental Biology\\University College London\\
\url{sihao.liu.18@ucl.ac.uk} \\
  \AND
  Augustine N. Mavor-Parker\\
  Research Department of Cell and Developmental Biology\\
  Centre for Artificial Intelligence\\ University College London
  \AND
  Caswell Barry\\
  Research Department of Cell and Developmental Biology\\ University College London}

\begin{document}

\maketitle

\begin{abstract}
We aimed to explore the capability of deep learning to approximate the function instantiated by biological neural circuits - the functional connectome. Using deep neural networks, we performed supervised learning with firing rate observations drawn from synthetically constructed neural circuits, as well as from an empirically supported Boundary Vector Cell–Place Cell network. The performance of trained networks was quantified using a range of criteria and tasks. Our results show that deep neural networks were able to capture the computations performed by synthetic biological networks with high accuracy, and were highly data efficient and robust to biological plasticity. We show that trained deep neural networks are able to perform zero-shot generalisation in novel environments, and allows for a wealth of tasks such as decoding the animal's location in space with high accuracy. Our study reveals a novel and promising direction in systems neuroscience, and can be expanded upon with a multitude of downstream applications, for example, goal-directed reinforcement learning.

\end{abstract}

\section{Introduction}
Artificial deep neural networks (DNNs) \cite{goodfellow2016deep} have been sucessfully applied to a range of domains including image recognition \cite{gu2018recent}, speech and language processing \cite{young2018recent} and reinforcement learning \cite{li2017deep}. Despite these success stories, the systems neuroscience community has been skeptical about deep learning being the solution to understanding intelligence in the brain. This is because of the simplification of individual neurons in these models as well as the biological plausibility of the back-propagation training algorithm \cite{rumelhart1986learning}. Nevertheless, DNNs (ReLU networks) are powerful tools as approximators \cite{hornik1989multilayer} despite abstracting away details from biology. \\

We explore how the architectures, including widths and depths of an arbitrary neuronal circuit corresponds to a DNN. We begin by constructing a rate network with additional features corresponding to sensory processing in the brain---which we call an \textit{Approximate Biological Neural Network} (ABNN)---and generate synthetic input and output patterns using simulated spike trains. We fit two classes of DNNs, Multi-Layer Perceptrons \cite{rosenblatt1958perceptron} (MLPs) and Long Short-Term Memory \cite{hochreiter1997long} networks (LSTMs) with various numbers of depths and widths are fitted to learn these patterns. We examine the data efficiency the approximating DNNs and how they generalise when the ABNN undergoes small plasticity changes. After, we simulate an experimentally supported rate network in the mammalian hippocampal formation---the Boundary Vector Cell model \cite{barry2007learning}---and fit the DNNs to the firing rates of the downstream place cells. Further, we demonstrate that the learned DNNs are able to perform a range of tasks related to the hippocampal formation, such as decoding the trajectory of an agent in an environment.\\
A DNN substituted neuronal circuit is useful as it allows fast and non-invasive experimental neuroscience research. We hope this preliminary work give rise to further research in complicated, hierarchical models of biological neural circuits, as well as applying the approaches developed to \textit{in vivo} experimental data rather than just theoretical models of neural circuits.

\subsection{Related Works}
Existing works focuses on modelling single the behaviour of single neurons---Beniaguev et al. \cite{beniaguev2021single} demonstrated that a 3D-reconstructed rate layer 5 cortical pyramidal neuron with NMDA receptors can be best captured by a ReLU network with 7 hidden layers and 128 hidden units each. Removing NMDA receptors allows the neuron to be learned by a much smaller deep network with a single hidden layer and 128 units. Moldwin et al. \cite{moldwin2020perceptron} implemented the perceptron learning algorithm \cite{rosenblatt1958perceptron} on a biophysical model of a simulated layer 5 pyramidal cell using the NEURON \cite{hines2001neuron} software package and found that the biophysical perceptron was able to achieved results comparable to a traditional perceptron model. Wang et al. \cite{wang2022predicting} successfully predicted the spike features as well as the dropping intervals in 9 simulated Hodgkin-Huxley neurons with different ion channel settings using a DNN trained on spike voltage data, which they dubbed Feature Prediction Module. 
\section{Methods}
Our experiments as structured as follows. First, we construct a series of deep learning models, that we call Approximate Biological Neural Networks (ABNNs), which contain features that make them somewhat more biologically plausible. Then we systematically study the ability of a separate neural network to learn to perform the same mapping as the ABNNs. After, we repeat this procedure for a more biologically plausible model of circuits in the hippocampal formation. 
\subsection{Learning an Approximate Biological Neural Network}
\begin{figure}
    \centering
    \includegraphics[scale=0.25]{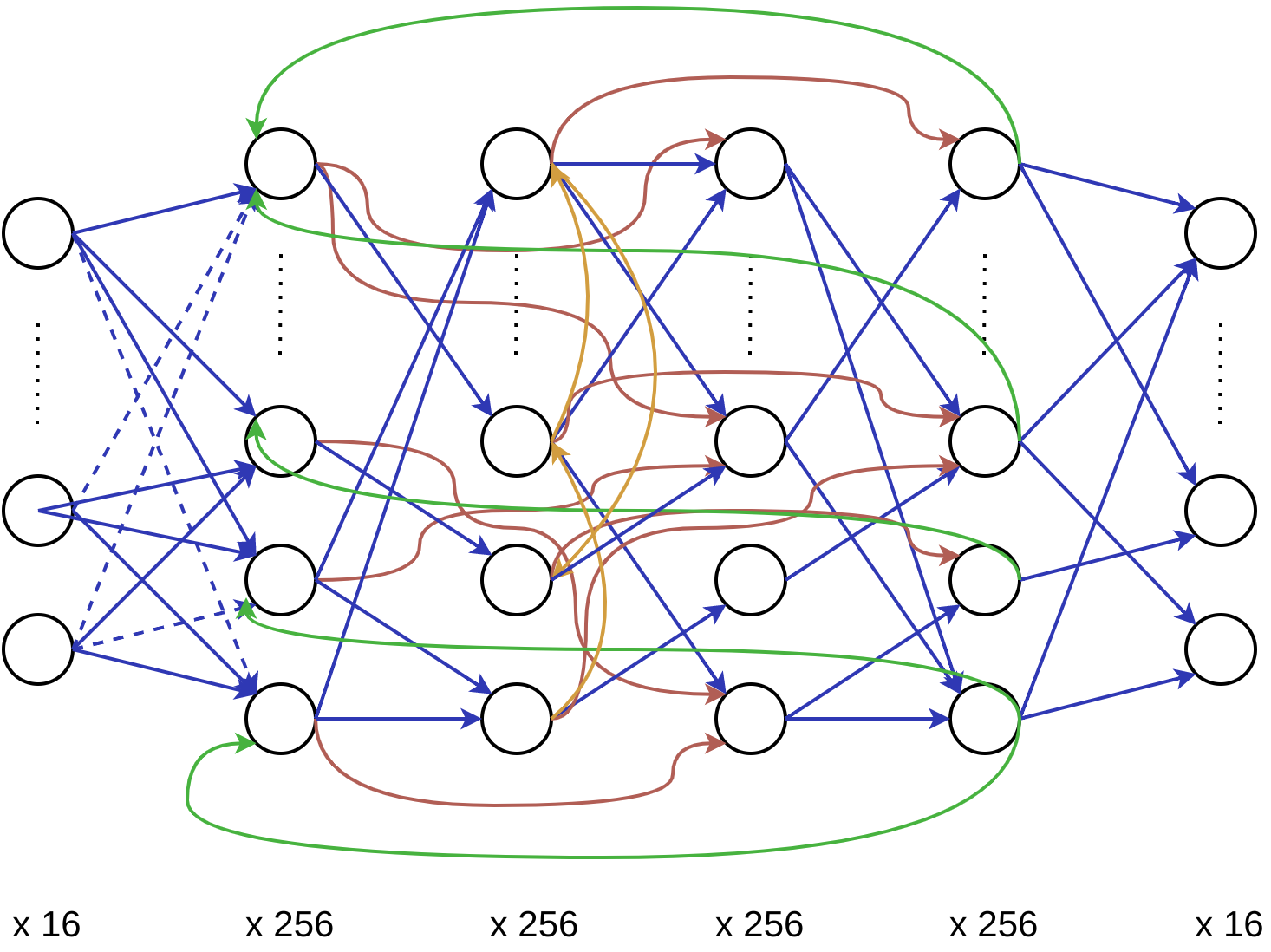}
    \caption{The structure of a full complex arbitrary neural circuit. Left is a simplified diagram showing information flow and right represents detailed connections between each layer, where (A) blue lines represent feed-forward connections; (B) dotted lines indicate weights dropped to zero to comply with sparsity and is not shown for subsequent layers to avoid cluttering; (C) red curves indicates skip connections between layers 1 and 3 as well as 2 and 4; (D) amber curves indicate lateral inhibition in layer 2; (E) green curves indicate feed-back connections from layer 4 to layer 1. The number of neurons in each layer is shown in the middle and not to scale in the right diagram.}
    \label{fig:approximate_bnn_schematic}
\end{figure}

The base structure of our ABNN resembles a MLP with $16$ input neurons, $16$ output neurons, and $4$ layers of $256$ hidden units each. On top of this scaffold we consider augmenting the following network properties.
\begin{enumerate}
    \item Variations between neurons. Neurons tend to have different biophysical properties, such as axonal length and membrane conductivity \cite{prescott2008biophysical}, that result in different firing behaviours. We model these biophysical difference by assigning each neuron in each layer a distinct transfer function \cite{tutorial2005transfer}. These are sampled randomly from ReLU, ELU, SiLU, CELU, Sigmoidal, Tanh and Leaky ReLU with leak parameters $0.1$, $0.2$ and $0.3$.
    \item Excitatory and inhibitory synapses. We consider the fact that various types of neurotransmitters govern whether the activation of a pre-synaptic neuron invoke excitatory or inhibitory responses in the post-synaptic neuron \cite{petroff2002book}. The weights between neuron $i$ in the previous layer and neuron $j$ in the next layer is initialised using a unit Gaussian: $w_{ij} \sim \mathcal{N}(0,1) \forall i,j$.
    \item Feed-forward connections that resemble the hierarchical architectures of mammalian neocortex \cite{byrne2013introduction} \cite{callaway2004feedforward}. For example, an ABNN with feedforward connections has 16 input neurons, projecting sequentially to 4 hidden layers of 256 neurons each, and to an output layer of 16 layers.
    \item Skip connections \cite{im2019skip}. Additional weights are initialised to downstream layers using the same schemes as above
    \item Feed-back connections \cite{byrne2013introduction}. We allow the neurons in downstream layers to project back to upstream layers, with the same weight initialisation scheme. This correspond to the observation of large number of recurrent connections in the brain.
    \item Sparsity. Real neurons in the brain are sparsely connected \cite{herculano2009human}. Each connection is randomly dropped with probability $0.5$. This feature is added based on the fact that neuronal connections in the brain are highly sparse.
    \item Lateral inhibitions \cite{byrne2013introduction}. We additionally allow each neuron $i$ in a hidden layer to connect to every other neuron in the same layer, with negative weight:
    \begin{equation}
        w_{ij} \sim \mathcal{N}(-1 + \frac{|i-j|}{N}, 1) \qquad \forall i, j \quad i \neq j
        \end{equation}
        Where $N$ is the total number of neurons in that layer. That is, any central neuron inhibits its neighbours. This feature follows from the observation that lateral inhibition is ubiquitous in early sensory areas.
    \end{enumerate}
Figure \ref{fig:approximate_bnn_schematic} shows an illustration of an ABNN architecture with the different design choices described above being color coded.

\paragraph{Synthetic Biological Data}
To mimic the information processing in a biological neural network, we use the number of spikes from each neuron as the input pattern to our ABNN described below. We assume the number of spikes encode the value of the external stimulus (\textit{rate coding}). To produce variable spike input patterns, spike trains are simulated from a point process with Gamma-distributed inter-spike intervals \cite{kuffler1957maintained}, which mimics the refractory period of real neuron firings \cite{mccormick2014membrane}. Specifically, the inter-spike intervals $t_i$ for each neuron $i$ in the input layer to follow a Gamma distribution:
\begin{equation}
    t_i \sim \Gamma(\alpha_i, \beta_i) \qquad \forall i
\end{equation}
Where the mean firing rate for each input neuron $i$ has a mean firing rate $r_i$, sampled from a Gaussian:
\begin{equation}
    \mathbf{E}(t_i) = \frac{\alpha_i}{\beta_i} = r_i \sim \mathcal{N}(10, 1) \qquad \forall i
\end{equation}
The simulated spike timings for each neuron are then time-binned to produce input patterns. Thus, each input pattern consists of vectors of spike counts produced by $16$ neurons in 50 time steps. These patterns are fed into the ABNN to producing ``raw'' output patterns from each of the 16 output neurons. A small Gaussian noise ($\sigma=0.01$) is added to the raw output patterns to model neural noise in the information passing process \cite{ma2006bayesian}, which are then normalised (Z-scored) to use as training labels.

\paragraph{An Assortment of Approximate Biological Neural Networks}
We construct 4 ABNNs, each with increasing complexity, following the properties described. These are summarised in Table \ref{abnn_table} For each ABNN, we use the same input neuron patterns, and generated $10,000$ pairs of input-output patterns as training set, $1,000$ as the validation set and $1,000$ as test set.
\begin{table}[H]
  \caption{Summary of structures or properties included with each ABNN.}
  \label{abnn_table}
  \centering
\begin{tabular}{c|ccllllll}
& Var. & Ex. \& in. & FF & Spar. & Skip & FB & LI & Noise \\ \hline
\textbf{Feed-forward ABNN} & \checkmark & \checkmark & \checkmark & \checkmark & & & & \checkmark \\
\textbf{Skip ABNN }        & \checkmark & \checkmark & \checkmark & \checkmark & \checkmark & & & \checkmark \\
\textbf{Feed-back ABNN}    & \checkmark & \checkmark & \checkmark & \checkmark & & \checkmark & & \checkmark \\
\textbf{Complex ABNN}      & \checkmark & \checkmark & \checkmark & \checkmark & \checkmark & \checkmark & \checkmark & \checkmark
% \bottomrule
\end{tabular}

\end{table}

\paragraph{Training Procedure}
To explore how we can exploit the power of deep learning, we trained various multi-layer perceptrons with varying width (number of hidden units) and depth (number of hidden layers). For the feed-back and complex ABNN, we also trained recurrent neural networks with an additional LSTM layer followed by varying numbers of linear layers. This follows from the inclusion of recurrent architectures in the ABNNs, which induces temporal dependency between trials. The LSTM layer has the same number of hidden units as the rest of the network.  All deep neural networks have ReLU non-linearities and are trained with a mean squared error loss function. For all ABNN experiments, we use the Adam optimiser \cite{kingma2014adam} with learning rate $0.001$ and batch size $200$, unless stated otherwise. Parameter search was run with a range of hidden layers $[1, 2, 3, 4, 5, 6 ,7, 8]$ and hidden units $[2, 4, 8, 16, 32, 64, 128, 256]$. The best combination of hyper-parameters is chosen by picking the set of parameters that minimised the mean final MSE loss on the test set. These networks are then trained on the training set, and the mean final MSE losses on the test set are reported. These are produced by averaging the MSE loss after training for 50 epochs over 30 repeats to ensure reliability. 
\paragraph{Data Efficiency, Generalisation and Plasticity}
We performed additional experiments on the complex ABNN and the corresponding trained DNNs. First, we explore the data efficiency of learning networks: DNNs are trained for up $1,000$ epochs using a subset $n$ of the $10,000$ training data vectors generated by the complex ABNN, drawn randomly without replacement and their performance were via MSE loss when predicting on the full, $1,000$ data vectors test set. The number of data vectors drawn are $n = 1, 5, 10, 50, 100, 200, 400, 800, 1600, 3200, 6400$. Secondly, we explore how the trained DNNs perform when plasticity changes occur in the complex ABNN. We model change by injecting a Gaussian random variable to each established connections:
\begin{equation}
    w_{ij} = w_{ij} + \delta w_{ij}, \qquad \delta w_{ij} \sim \mathcal{N}(0, \sigma^2) \qquad \forall i,j: w_{ij} \neq 0
\end{equation}
Where neuron $i$ and $j$ are connected. Another view of the weight change experiment is how well the DNNs can cope with random noises in the ABNN. The original input patterns are then passed through the altered ABNN and normalised again. The architecture and transfer function of the neurons are kept unchanged. $\sigma$ is systematically discretised in a logarithmic scale from $e^{-7.0}$ to $e^{4.0}$ and the performance of the trained DNNs are compared with randomly initialised ones. The experiment is repeated 30 times and the mean and standard deviation are taken to improve reliability.

\subsection{Learning the Boundary Vector Cell model}
\begin{figure}[H]
    \centering
    \includegraphics[scale=0.12]{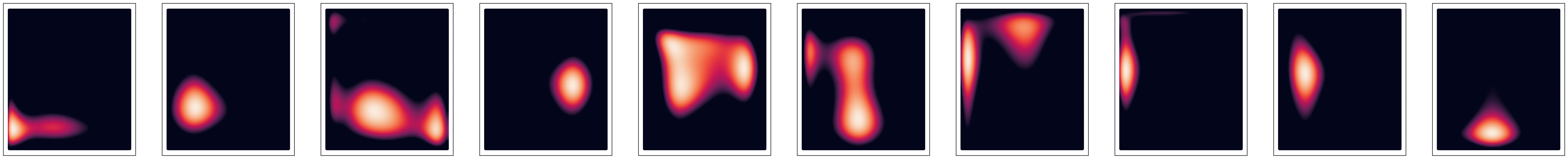}
    \caption{Bird's eye view of the firing fields of 10 Place Cells in a simulated square environment.}
    \label{fig:true_pc_firing_field}
\end{figure}

Next, we turn to an empirically supported model of the cortical circuit, the Boundary Vector Cell model\cite{barry2007learning}. Boundary vector cells are crucial for an animal's ability to self-localise in an environment. Analogous to the ABNN experiments, we run parameter search and train DNNs on the input-output pairs of the neural circuit in a supervised manner, using randomly sampled locations in an environment. We measure the generalisability of our approximations by making predictions in novel unseen environments and comparing the predictions to the ground truth from the BVC model.
\paragraph{Simulated Boundary Vector Cells and Place Cells}
Boundary vector cells \cite{lever2009boundary} are a type of pyramidal neurons found in the subiculum and entorhinal cortex of mammalian hippocampal formation. They respond to the presence of boundaries in an environment at a preferred distance $d_i$ and preferred bearing $\phi_i$ (in radians). The firing contribution $g_i$ of a small segment of the boundary subtending a small angle $\delta \theta$, with distance $r$ and bearing $\theta$ can be described with a tuning function.
\begin{equation}
g_i(r, \theta) \propto \frac{\exp[-(r-d_i)^2/(2\sigma_{\text{rad}}^2(d_i))]}{\sqrt{2 \pi \sigma_{\text{rad}}^2(d_i)}} \times \frac{\exp[-(\theta-\phi_i)^2/2\sigma_{\text{ang}}^2]}{\sqrt{2 \pi \sigma_{\text{ang}}^2}}
\label{eq: bvc_segment_firing}
\end{equation}
The overall firing rate of a boundary vector cell $f_i$ at a location $\mathbf{x}$ is obtained by integrating over all over all angles:
\begin{equation}
f_i(\mathbf{x}) = \int_0^{2 \pi} g_i(r, \theta) \delta \theta
\label{eq: bvc_global_firing}
\end{equation}
Boundary vector cells project to place cells \cite{o1971hippocampus}, another type of pyramidal neuron in the hippocampus proper. In \cite{lever2009boundary}'s model, the firing rate of a place cell $F_j$ is a thresholded sum of the firing rate of upstream boundary vector cells:
\begin{equation}
F_j(\mathbf{x}) = \left [ A \sum_{i=1}^{N} w_{ij} f_i(\mathbf{x}) - T_j\right ]_+
\label{eq: pc_global_firing}
\end{equation}
Where $T_j$ is the threshold and $[\cdot]_+$ denotes the non-linear rectifier. We simulate $100$ boundary vector cells and $10$ place cells. Each boundary vector cell $i$ has preferred distance $d_i$ and angle $\phi_i$ sampled from continuous uniform distribution:
\begin{equation}
    d_i \sim \mathcal{U}(0, 3200) \qquad \phi_i \sim \mathcal{U}(0, 2\pi) \qquad \forall i = 1, ..., 100
\end{equation}
Each place cell $j$ is randomly connected to $N=15$ BVCs. The connection weights are sampled according to a normal distribution:
\begin{equation}
    w_{ij} \sim \mathcal{N}(1, 1)
\end{equation}
The multiplier is chosen to be $A=10,000$ and the threshold $T_j$ is capped at $80\%$ of the firing capacity of that place cell \cite{hartley2000modeling}. The rate map of the resulting place cells are illustrated in Figure \ref{fig:true_pc_firing_field}.
We sample random locations in a $3,200 \times 3,200$ mm square environment, and simulate the firing rates of the boundary vector cells and place cells as input and output patterns, respectively. $10,000$ pairs are sampled as the training set, $1,000$ pairs for the validation set and $1,000$ pairs for the test set.

\paragraph{Training Deep Networks}
We perform hyper-parameter search on a range of multi-layer perceptrons with hidden linear layers $[1, 2, 3, 4, 5, 6, 7, 8]$ with a range of hidden units $[1, 2, 4, 8, 16, 32, 64, 128, 256]$, mapping from an $100$ input neuron firing rates to $10$ output neuron firing rates. We trained these networks on the validation set for 100 epochs and find the network with the best combination of hyperparameters over 30 repeats. This network is then trained with the training set data and tested using vairous tasks. In all training, mean squared error is used as the loss function and Adam \cite{kingma2014adam} with learning rate $0.001$ is used as the optimiser.

\begin{figure}[H]
    \centering
    \includegraphics[scale=0.13]{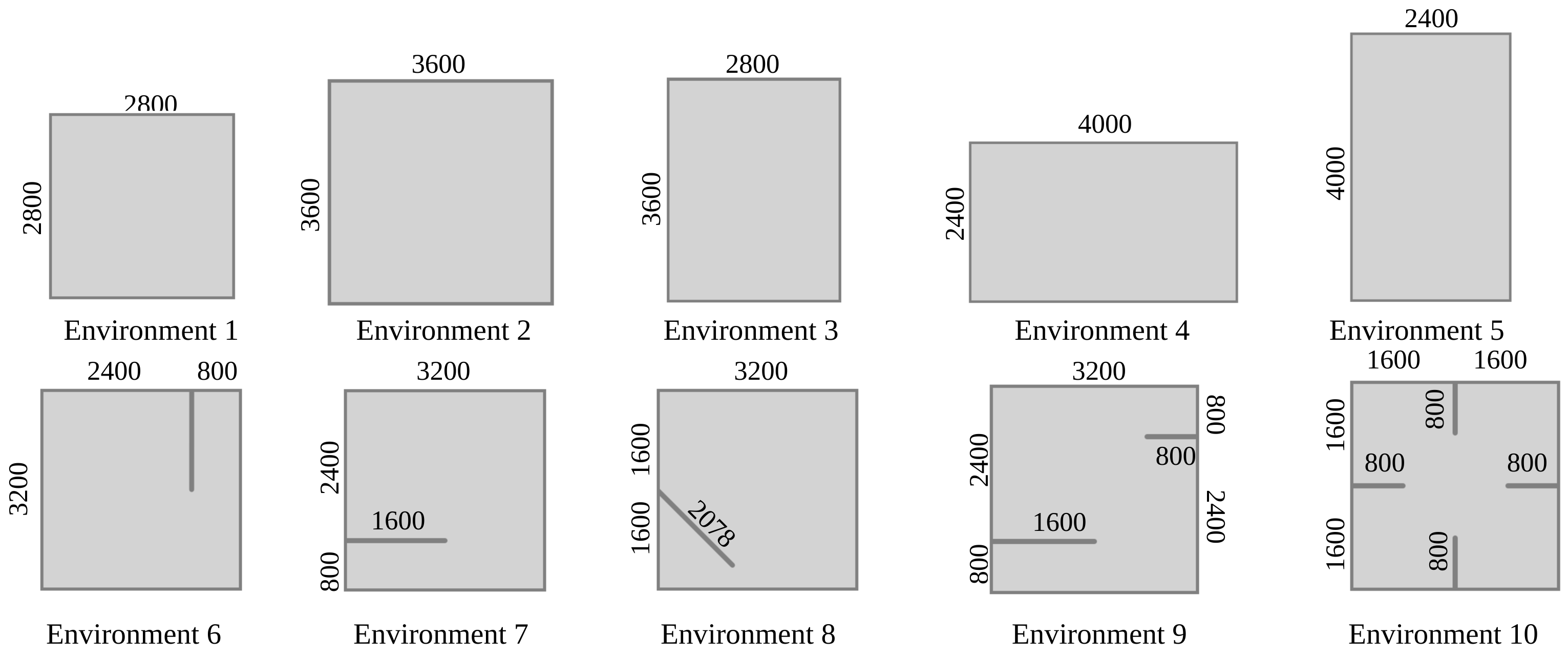}
    \caption{Diagrams showing 10 unseen environments.}
    \label{fig:10_unseen_envs}
\end{figure}

\paragraph{Generalisation in Unseen Environments}
We test the generalisation performance of the trained DNN in unseen environments. The $10$ place fields are predicted on two groups of environments. In the first group, we stretch and compress the length and width of the rectangular environment: environments with shape $2,800 \times 2,800$, $3,600 \times 3,600$, $2,800 \times 3,600$, $4,000 \times 2,400$, $2,400 \times 4,000$. In the second group, we insert random straight barriers into the environments. Note that this alters the firing fields of boundary vector cells and thus the connected place cells. These are presented as Environments 1-5 and 6-10 in Figure \ref{fig:10_unseen_envs}.
In both groups, we evaluate the performance of firing field prediction using the mean squared error loss and the structural similarity index (SSIM) \cite{ponomarenko2018structural}. SSIM was designed to compare changes in quality of digital images due to compression. It has the benefit of being insensitive to shifts and translations, making it a good metric for evaluating the shape of the predicted firing fields.It is made up of three components: luminance $l(\mathbf{x,y})$, contrast $c(\mathbf{x,y})$ and structure $s(\mathbf{x,y})$:
\begin{equation}
    l(\mathbf{x,y}) = \frac{2\mu_x \mu_y + c_1}{\mu_x^2 + \mu_y^2 + c_1} \qquad c(\mathbf{x,y}) = \frac{2 \sigma_x \sigma_y + c_2}{\sigma_x^2 + \sigma_y^2 + c_2} \qquad s(\mathbf{x,y}) = \frac{\sigma_{xy} + c_3}{\sigma_x \sigma_y + c_3}
\label{eq: ssim_1}
\end{equation}
Where $\mu_\cdot$, $\sigma_\cdot$, $\sigma_{xy}$ are respectively the mean, bias-corrected standard deviation and bias-corrected covariance across all pixels values in $\mathbf{x}$ and $\mathbf{y}$. We take $c_1 = (0.01L)^2, c_2 = (0.03L)^2$, and $c_3 = 0.5c_2$ where $L$ is the maximum place cell firing rate in each firing field. The resulting SSIM is the exponentially weighted product of the three attributes from \ref{eq: ssim_1}:
\begin{equation}
    \text{SSIM}(\mathbf{x,y}) = [l(\mathbf{x,y})]^\alpha \times [c(\mathbf{x,y})]^\beta \times [s(\mathbf{x,y})]^\gamma
\label{eq: ssim_2}
\end{equation}
Where we consider the three attributes weighted equally: $\alpha = \beta = \gamma = 1$.

\paragraph{Bayesian Decoding of Agent Trajectory}
In this task, we test whether using the surrogate deep network, one can decode the trajectory of an agent. We consider an agent carrying the boundary vector cell--place cell neural circuit roaming freely in an unseen environment with barriers, following a Brownian motion model, inspired by the \textit{RatInABox} package \cite{george2022ratinabox}. Given the recorded firing rates of boundary vector cells at each time-step, we use a Bayesian decoder to obtain a \textit{maximum a posteriori} trajectory of the agent at each time step, comparing the decoding quality with predicted firing rates of place cells, true firing rates of place cells, predicted rate maps of place cells and true rate maps of place cells. An additional preprocessing step is required to ensure the DNN predicted firing fields and firing rates over all time steps are non-negative, thus these data needs to be pre-processed by passing through a rectifier function.

\section{Results}
\subsection{Approximate Biological Neural Network}
\begin{figure}[H]
    \centering
    \includegraphics[scale=0.4]{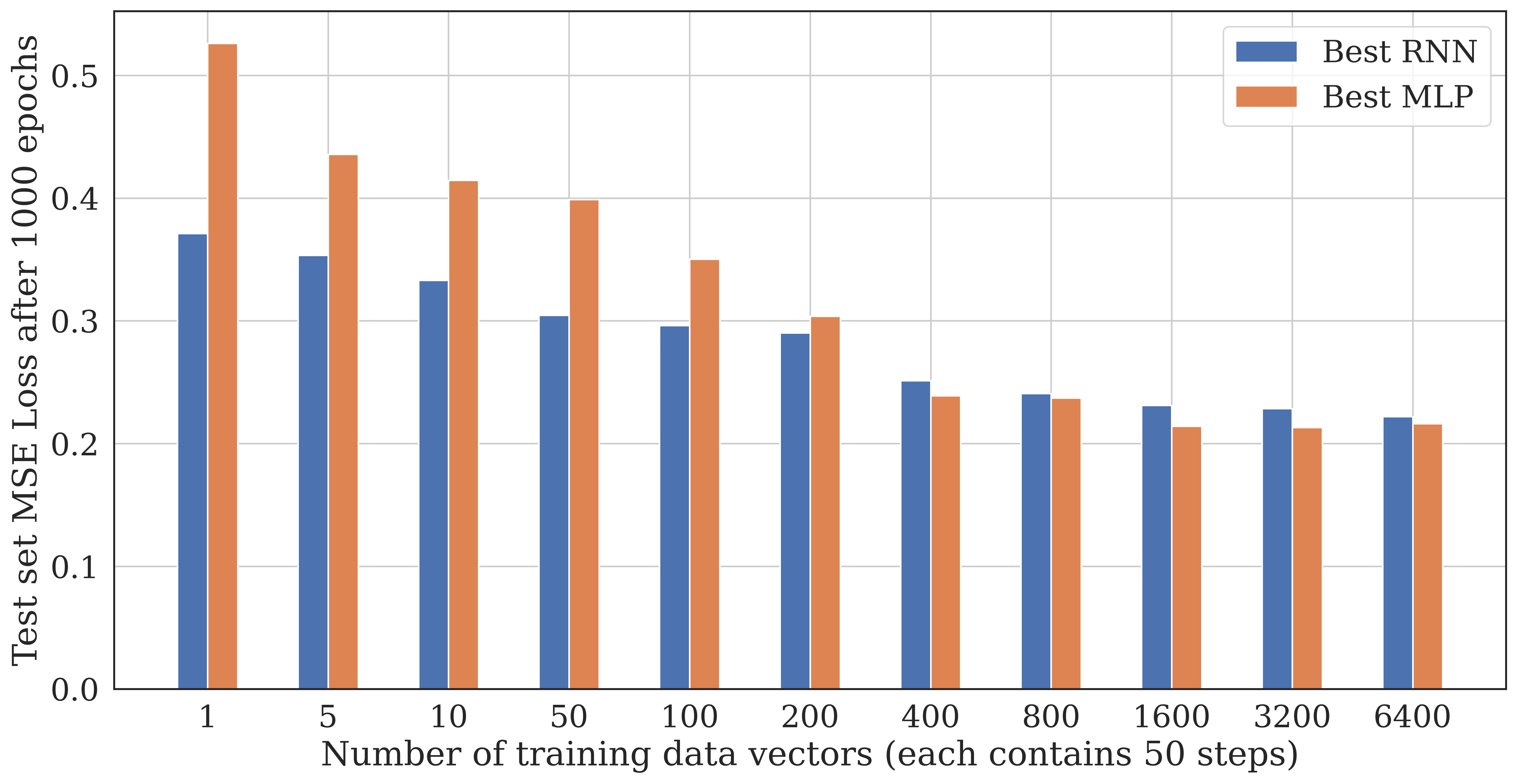}
    \caption{Best average MSE loss by each DNN achieved over $1,000$ epochs, with the number of training samples specified on the X-axis.}
    \label{fig:data_efficiency_comparison}
\end{figure}

\begin{figure}[H]
    \centering
    \includegraphics[scale=0.4]{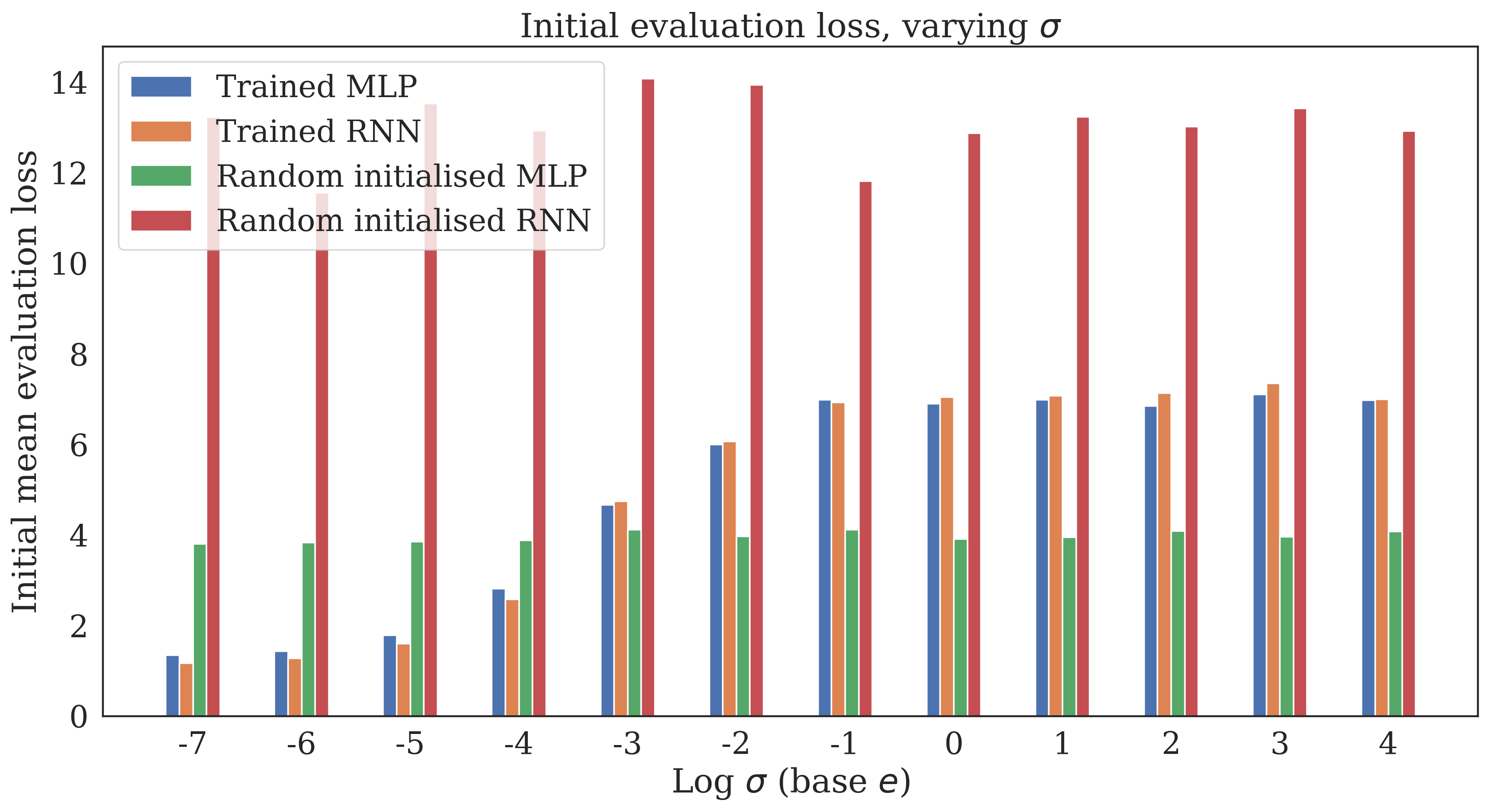}
    \caption{Average initial MSE loss (Y-axis) by: (blue) trained MLP; (orange) trained RNN; (green) a randomly initialised MLP and (red) a randomly initialised RNN over a range of noise $\sigma$ (X-axis).}
    \label{fig:plasticity_init_loss}
\end{figure}

An initial analysis of training set output distribution (before adding Gaussian noise) by each output neurons in the ABNNs shows that the majority attains the maximum empirical entropy--the firing rates of these neurons are different for each distinct input pattern--whereas few output neurons show unvarying outputs. This is because these neurons were assigned Sigmoidal or Tanh transfer functions so that the ouput is capped at $1$, $0$ or $-1$. Further, a Principal Component Analysis \cite{pearson1901liii}  shows the variance of the first few eigen-directions delay slowly ($2.34, 1.78, 1.41$ for the first 3 principal components in the complex ABNN output). We conclude that this an appropriate model of biological networks in the brain as a realistic model of neural circuits in the brain, as information is concentrated in a large subset of neurons, rather than a few.
\begin{table}[H]
\centering
\caption{Summary of parameter search and training result with each ABNN}
\begin{tabular}{cc|cccc}
ABNN type                       & DNN & layers & units & test MSE    & baseline                    \\ \hline
\textbf{Feed-forward ABNN  }    & MLP & 4             & 256          & 0.1099 ± 0.0060 & 0.7493                  \\ \hline
\textbf{Skip ABNN}              & MLP & 3             & 256          & 0.0881 ± 0.0088 & 0.8530                  \\ \hline
\textbf{Feed-back ABNN}         & MLP & 2             & 256          & 0.2678 ± 0.0063 & 0.9991 \\ 
                                & RNN & 3             & 256          & 0.2594 ± 0.0011 &                         \\ \hline
\textbf{Complex ABNN}           & MLP & 5             & 256          & 0.2472 ± 0.0052 & 0.7593 \\
                                & RNN & 8             & 256          & 0.2196 ± 0.0008 &                        
\end{tabular}
\label{table:abnn_exp}
\end{table}
Table \ref{table:abnn_exp} summarises the parameter search results, as well as the trained network performance on the test set. As a comparison, the baseline test set MSE is achieved by predicting the mean of each dimension. In all four approximate biological neural networks, the trained DNNs perform significantly better than the baseline, meaning the DNNs are able to capture some of the information processing properties of the biological counterparts. Varying the number of training data $n$ generated from the complex ABNN, the best test set MSE loss achieved by the MLP and RNN for training up to $1,000$ epochs are shown in Figure \ref{fig:data_efficiency_comparison}. Modelling plasticity changes in ABNN by injecting Gaussian noise to the connection weights, the test set loss for each value of standard deviation $\sigma$ is shown in Figure \ref{fig:plasticity_init_loss}.
As a comparison, randomly initialised, untrained DNNs of the same architectures are used for the baseline test set MSE loss. The mean is taken over 30 repeats. We then proceed perform transfer learning on the trained DNNs on the new data pairs, and see if the pre-training gives any advantage compared to random initialisation. The results are shown in the figures below
% \begin{figure}[H]
%      \centering
%      \includegraphics[scale=0.4]{images/retrain_generalisation.png}
%      \caption{Average test set MSE loss for transfer learning and training 4 DNNs over 30 epochs.}
%      \label{fig:plasticity_retrain_loss}
% \end{figure}
\begin{figure}[H]
     \centering
     \begin{subfigure}[b]{0.49\textwidth}
        \centering
        \includegraphics[width=\textwidth]{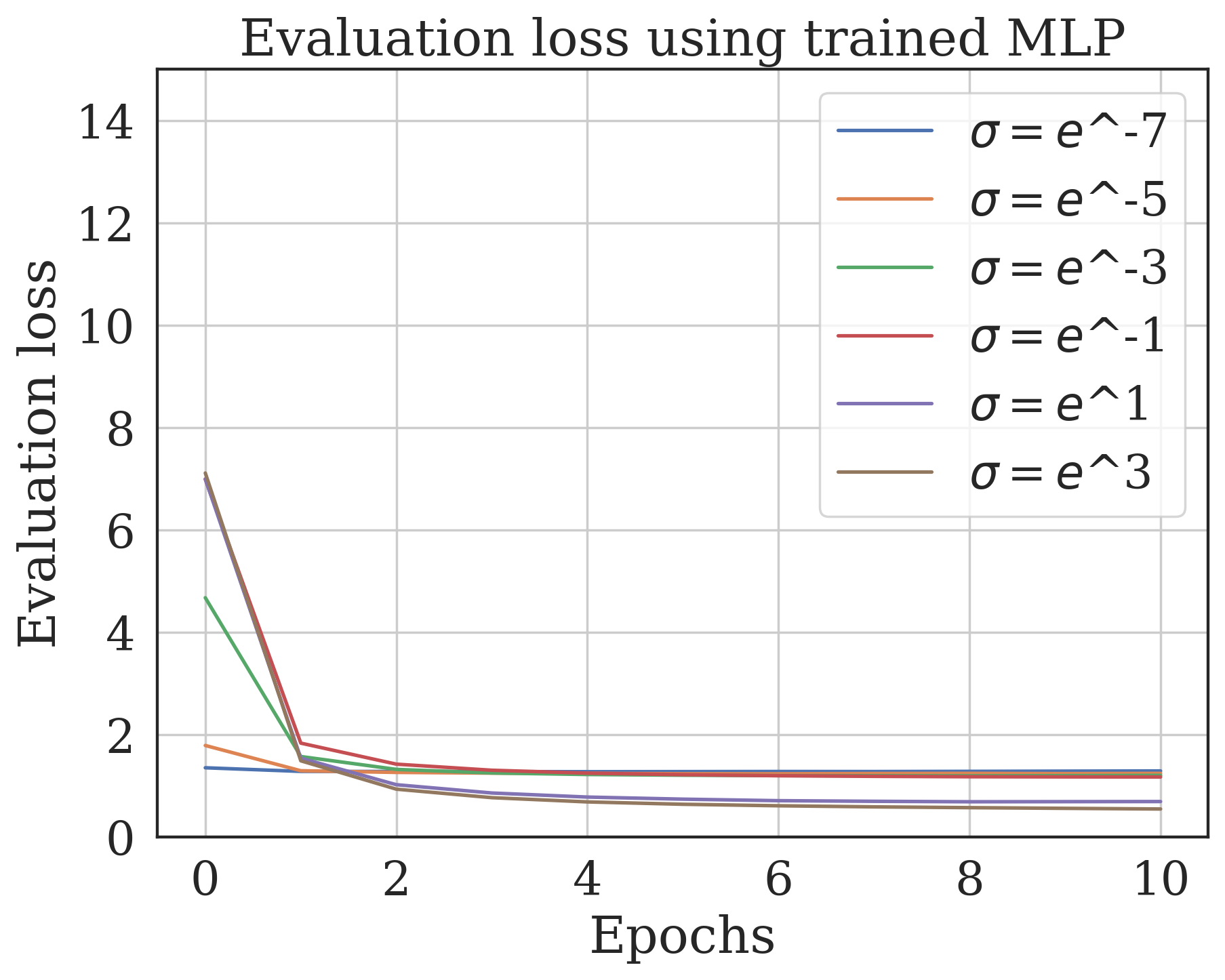}
        \label{fig:plasticity_retrain_loss_1}
    \end{subfigure}
    \begin{subfigure}[b]{0.49\textwidth}
        \centering
        \includegraphics[width=\textwidth]{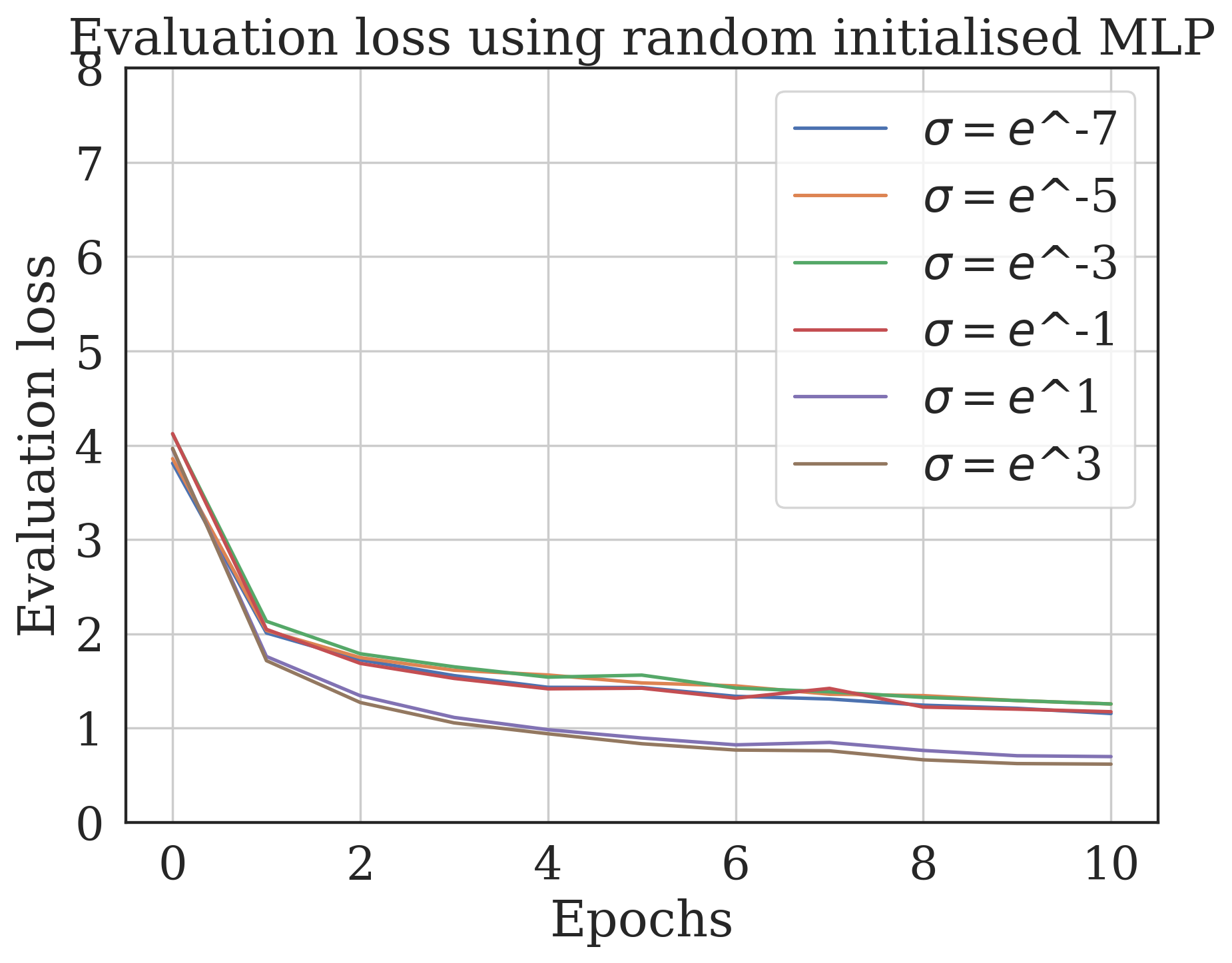}
        \label{fig:plasticity_retrain_loss_2}
    \end{subfigure}
    \begin{subfigure}[b]{0.49\textwidth}
        \centering
        \includegraphics[width=\textwidth]{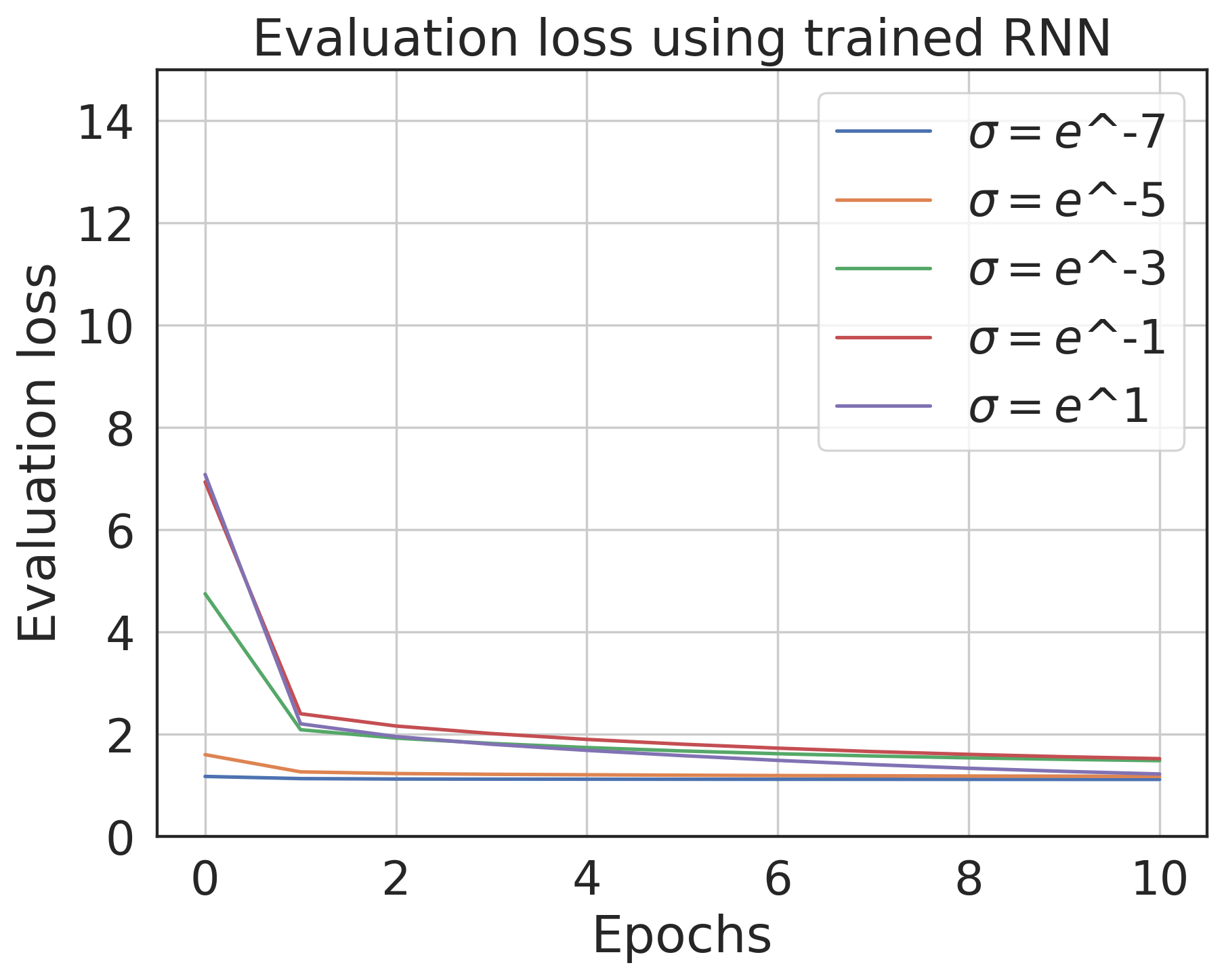}
        \label{fig:plasticity_retrain_loss_3}
    \end{subfigure}
    \begin{subfigure}[b]{0.49\textwidth}
        \centering
        \includegraphics[width=\textwidth]{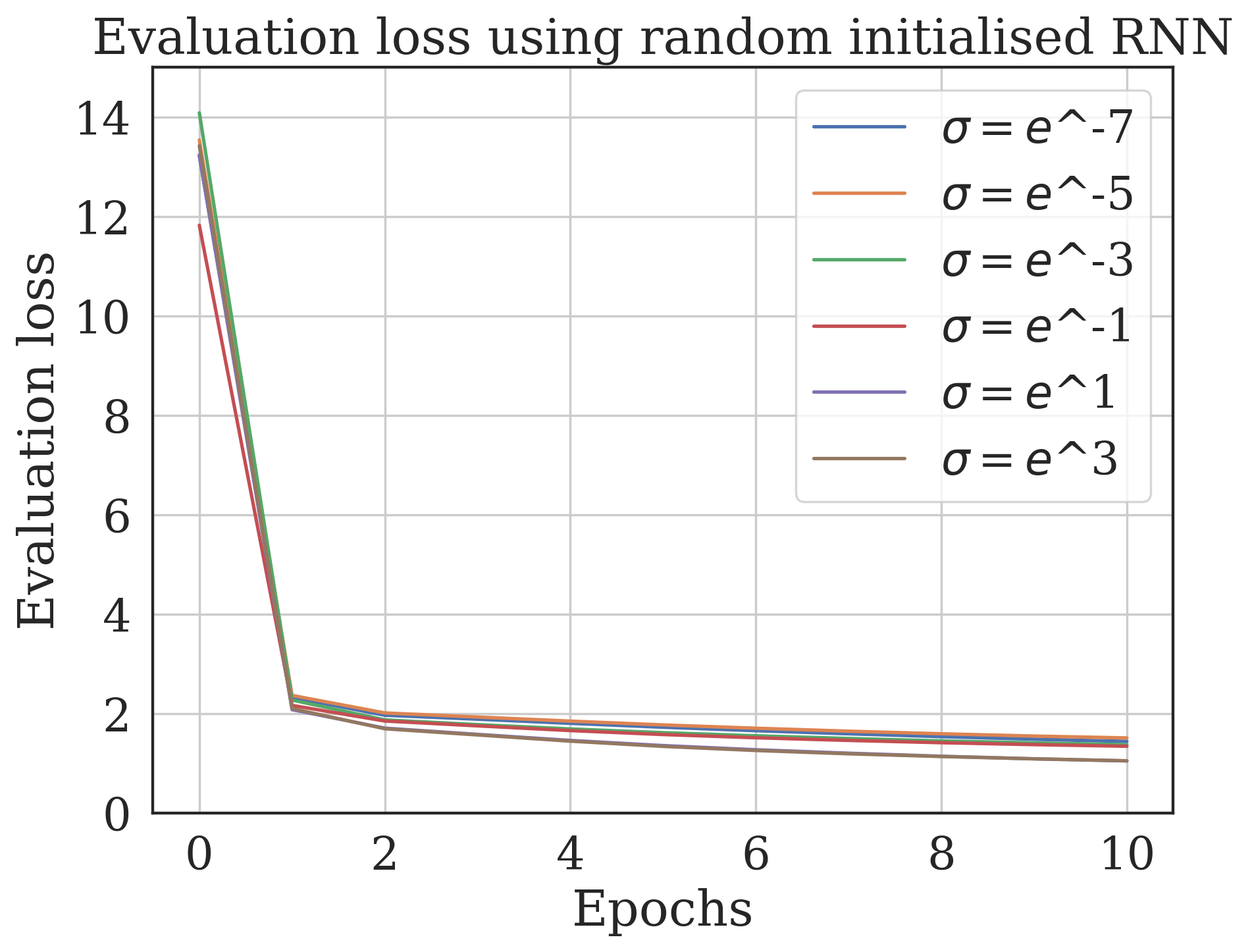}
        \label{fig:plasticity_retrain_loss_4}
    \end{subfigure}
\end{figure}

\subsection{Boundary Vector Cell Model}
\begin{figure}[H]
    \centering
    \includegraphics[scale=0.25]{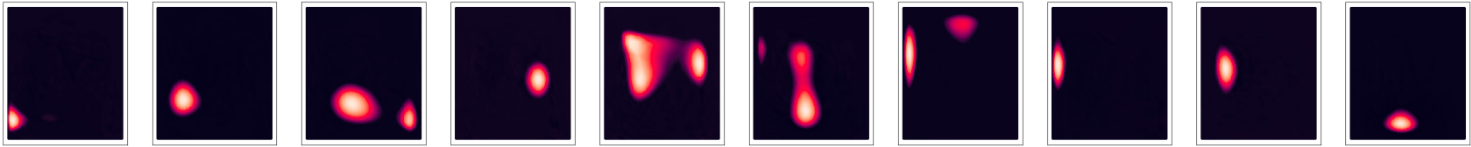}
    \caption{DNN predicted place fields of 10 simulated place cells in the environment.}
    \label{fig:predicted_pc_firing_field}
\end{figure}

\begin{figure}[H]
     \centering
     \begin{subfigure}[b]{0.49\textwidth}
        \centering
        \includegraphics[width=\textwidth]{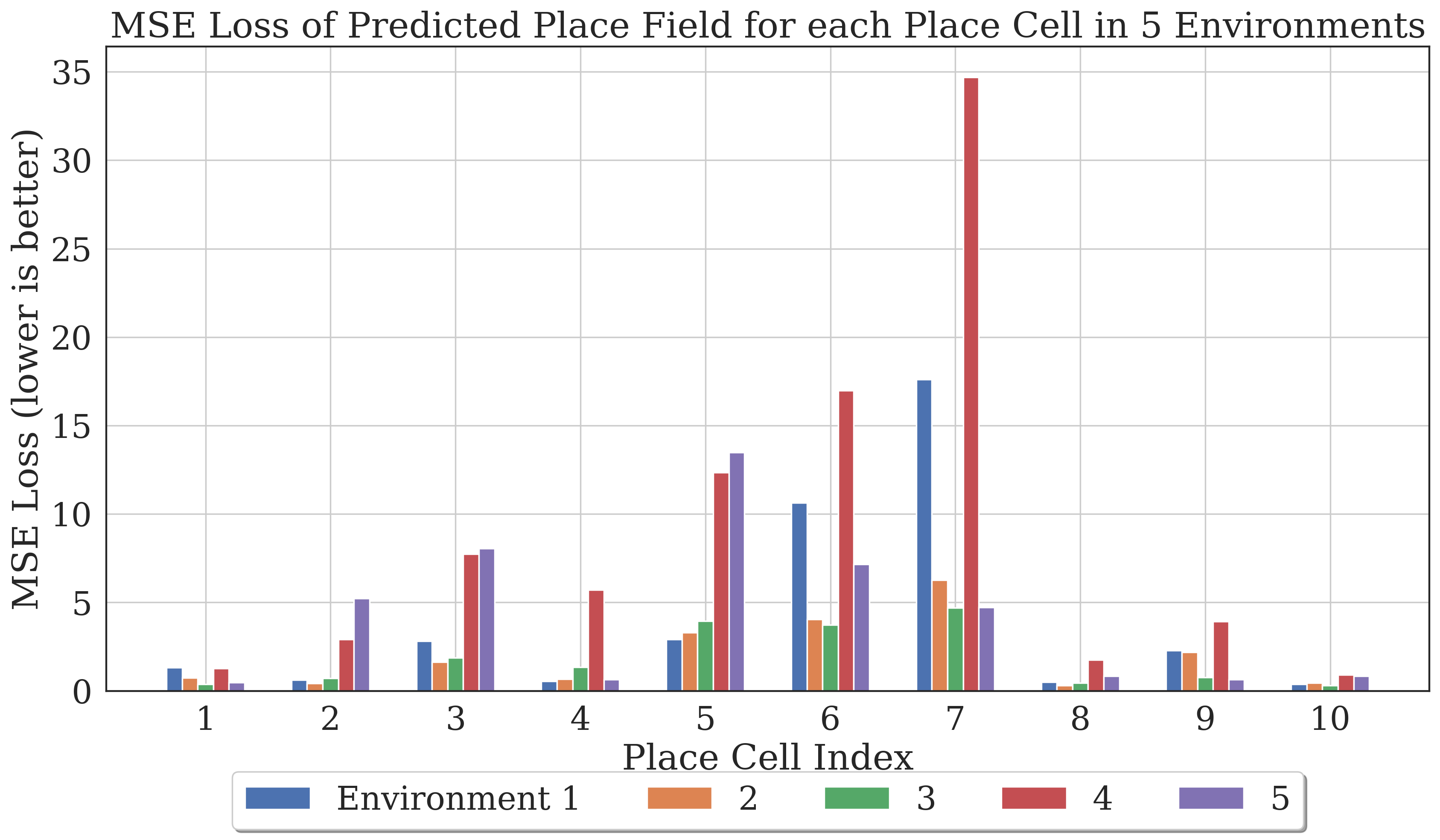}
        \caption{MSE for each predicted place field for 5 environments}
    \end{subfigure}
    \begin{subfigure}[b]{0.49\textwidth}
        \centering
        \includegraphics[width=\textwidth]{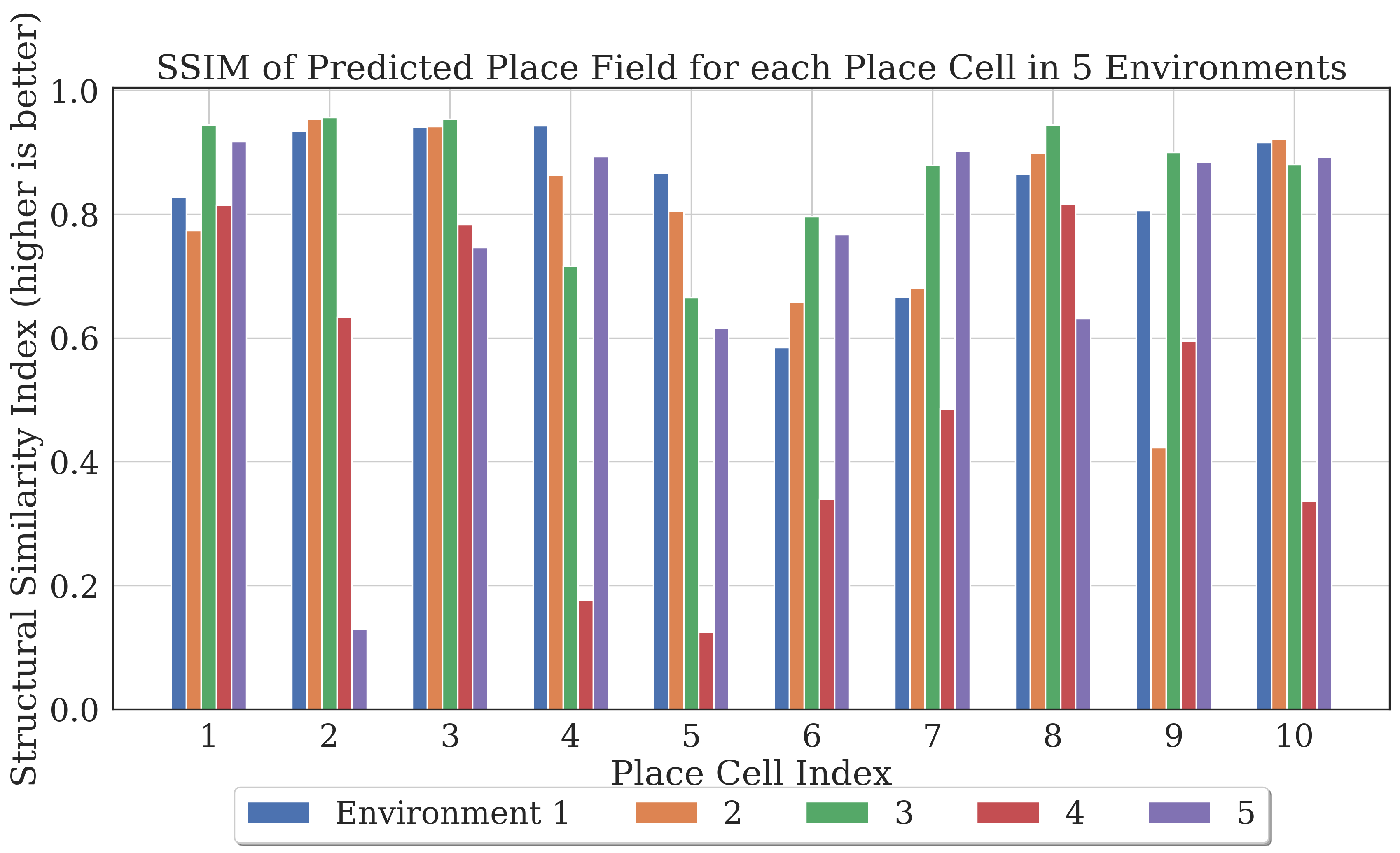}
        \caption{SSIM for each predicted place field for 5 environments}
    \end{subfigure}
    \caption{Evaluation of DNN prediction in Environments 1-5}
    \label{fig:eval_envs_1_5}
\end{figure}

\begin{figure}[H]
     \centering
     \begin{subfigure}[b]{0.49\textwidth}
        \centering
        \includegraphics[width=\textwidth]{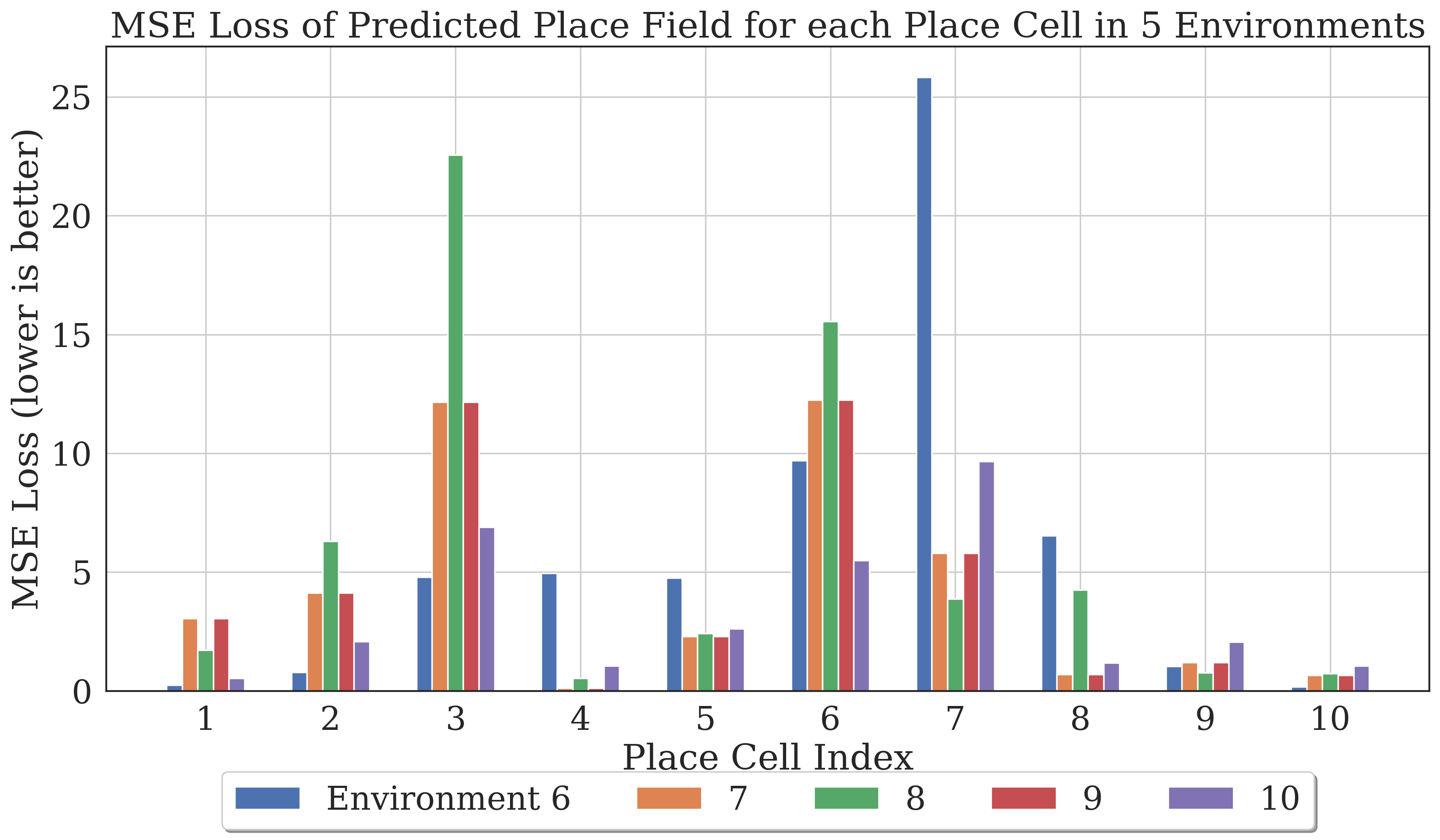}
        \caption{MSE for each predicted place field for 5 environments with barriers}
    \end{subfigure}
    \begin{subfigure}[b]{0.49\textwidth}
        \centering
        \includegraphics[width=\textwidth]{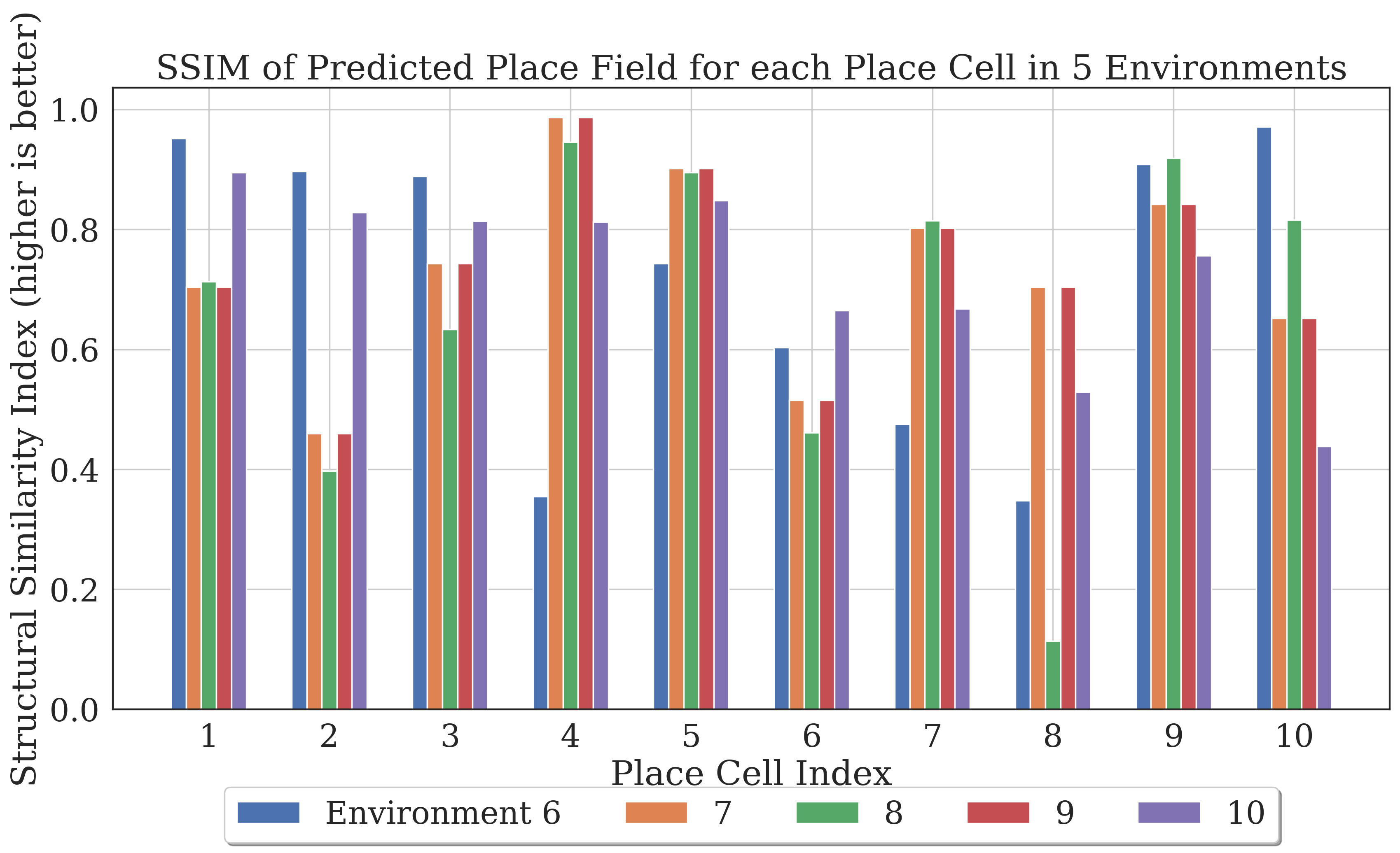}
        \caption{SSIM for each predicted place field for 5 environments with barriers}
    \end{subfigure}
    \caption{Evaluation of DNN prediction in Environments 6-10}
    \label{fig:eval_envs_6_10}
\end{figure}

\begin{figure}[H]
     \centering
     \includegraphics[scale=0.17]{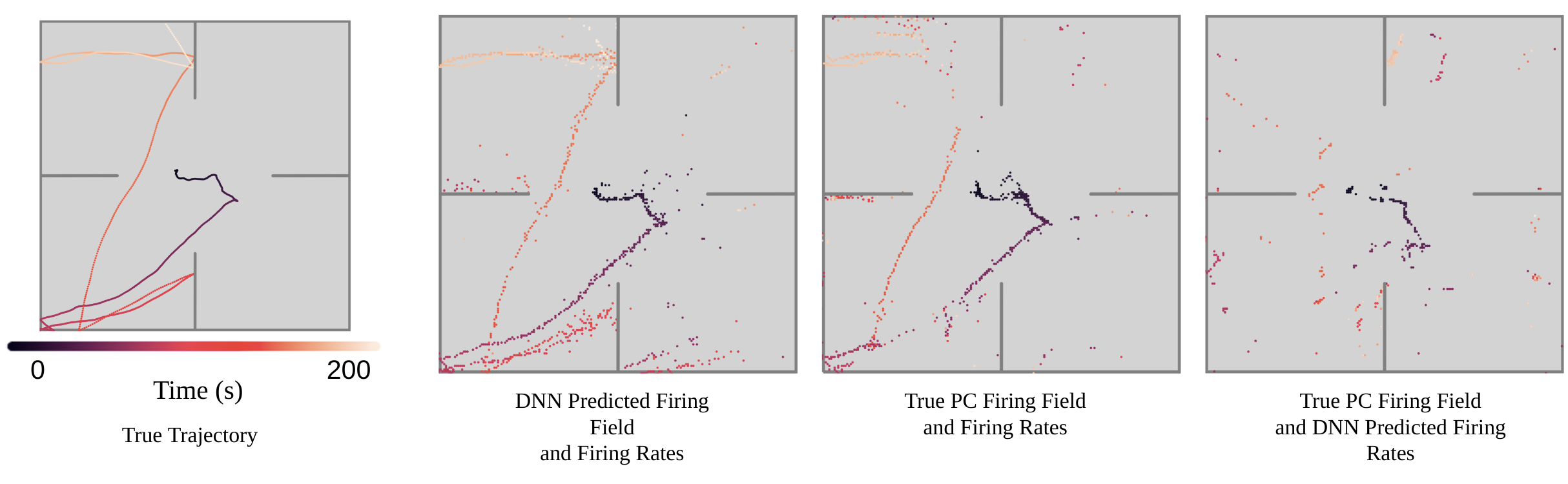}
     \caption{From left to right: true trajectory, decoded trajectories using (1) DNN predicted place cell rate maps and firing rates; (2) simulated place cell rate maps and firing rates; (3) DNN predicted firing rates and simulated place cell rate maps.}
     \label{fig:trajectories}
\end{figure}

The $10$ simulated place cells have the rate map as shown in Figure \ref{fig:true_pc_firing_field}. The parameter search found a multilayer-perceptron with $6$ hidden layers and $256$ hidden units yields the lowest average test set MSE loss over 30 repeats. Using this network, we discretised the square environment, sampled the BVC firing rates at each mesh point and predicted the rate map of the 10 place cells. The result is shown in Figure \ref{fig:predicted_pc_firing_field}, and to be compared with Figure \ref{fig:true_pc_firing_field}.
The trained DNN is then tested on the 10 unseen environments created above. When evaluated using both mean squared error and SSIM. The results are illustrated in Figure \ref{fig:eval_envs_1_5} and Figure \ref{fig:eval_envs_6_10}.
Inserting barriers into the environment changes the place fields drastically as the upstream boundary vector cells also fire in response to these barriers. Nevertheless, the trained DNN is able to predict the resulting place fields with good quality, as indicated by relatively low MSE loss ($<5$) and high SSIM ($>0.6$) in the majority of place cells. This shows the trained DNN was able to learn the mapping between boundary vector cell and the place cells firing rates, rather than simply remembering the place fields in the original environment. High structural similarity was also found for the majority of the place fields. An observation is that the prediction quality becomes poor when a barrier directly overlaps with the original place field, leading to a place field that disappears on one side of the barrier (e.g. place cell 3 in Environment 7) or scatters along the end of the environment (e.g. place cell 1 in Environment 7).\\
\newline
A random trajectory of an agent roaming in Environment 10 is shown in Figure \ref{fig:trajectories}. We used a Bayesian \textit{maximum a posteriori} rate decoder to decode the trajectory using simulated boundary vector cell firing rates and (1) DNN predicted place cell rate maps and firing rates; (2) simulated place cell rate maps and firing rates; (3) DNN predicted firing rates and simulated place cell rate maps. For three cases respectively, $88.8\%$, $46.6\%$ and $1.2\%$ of all locations were within $320$ mm ($10$ discretised grid points by Manhattan distance) of the true location.

\section{Discussion}
Our results found that in feed-back and complex ABNNs, training with recurrent neural network with a single LSTM layer generalises better than vanilla neural networks. We believe this is because the LSTM layer is able to capture temporal dynamics due to the inclusion of feed-back and lateral inhibition connections. Furthermore, it was found that during training, MLPS converged in both training and test set MSE loss after $100-200$ epochs and begins to over-fit shortly afterwards, whereas the LSTM losses decreasing asymptotically over the entirety of $1,000$ epochs. This shows that the MLP requires less computational resources while providing a predicting power not significantly worse than the RNN. In all four ABNNs, we found DNNs with $256$ hidden units and a similar number of hidden layers perform optimally, matching the ABNN counterpart.\\
\newline
We also see that supervised learning on the ABNN is extremely data efficient. The DNNs produced $32.2\%$ (for RNN) and $22.9\%$ (for MLP) higher loss when trained on only $200$ out of $10,000$ data samples compared to $10,000$. The loss differences are reduced to $9.7\%$ (for RNN) and insignificant (for MLP) when increasing the number of training samples to $1,600$ data samples reduces the loss differences to . This result aligns with the neuroscientific observations that our brains learn with high efficiency \cite{van2009efficiency}, and draws correspondence between trained DNNs and real brain networks. In the generalisation and plasticity experiment, we first notice that randomly initialised RNNs yield much higher loss than any other networks, including the randomly initialised MLPs. We suspect that this is due to the weight initialisation mechanism being different in the LSTM layer---resulting in a drastically different range of predicted output upon presenting the biologically interpretable input patterns. Secondly, we observe that the performances of trained networks are better than untrained networks up to $\sigma = e^{-4.0}$, equivalent to $1.8\%$ of the standard deviation of the originally initialised weights---it should be recognised that every synaptic connection in a neural circuit undergoes $1.8\%$ change is quite considerable. Finally, we see that the trained DNNs at $\sigma > e^{-2.0}$ performs worse than any randomly initialised MLP. We believe that this is due to a \textit{distributional shift} in the dataset---the weights in the ABNN underwent substantial change such that the output patterns have a completely different distribution from the original ones. The trained DNNs were fit to the original output patterns, hence producing high loss. This can be seen a performing \textit{domain adaptation}. When retrained to fit the plasticity-altered dataset, whilst the advantage of transfer learning can be seen in small $\sigma$, we found that the performance quickly becomes indistinguishable after the first few epochs compared to randomly initialised DNNs and the MSE loss converges to similar values.\\
\newline
Training the DNN using data pairs sampled from the environments using the Boundary Vector Cell model finds high prediction accuracy in the rate maps of all 10 place cells, which are visually indistinguishable. Testing the trained DNN in unseen environments (Environments 1-5) sees it generalises with relatively high accuracy in environments with similar boundary length as the environment it was trained on: shrinking or expanding the environment by $1/7$ (Environment 1 and 2) sees the majority of place cell firing fields predicted with low ($< 2.0$) MSE and high ($>0.6$) SSIM. Changing the aspect ratio slightly (Environment 3) does not seem to affect the prediction quality by a large margin. By changing the environmental size and shape more drastically, the DNN prediction quality is further reduced (Environment 4 and 5). Nevertheless, the DNNs seems to yield good results for place cells whose firing fields are near the border of the environment (place cell indices 1 and 8).\\
\newline
Inserting barriers into the environment changes the place fields drastically as the upstream boundary vector cells also fire in response to these barriers. Nevertheless, the trained DNN is able to predict the resulting place fields with good quality, as indicated by relatively low MSE loss ($<5$) and high SSIM ($>0.6$) in the majority of place cells. This shows the trained DNN was able to learn the mapping between boundary vector cell and the place cells firing rates, rather than simply remembering the place fields in the original environment. High structural similarity was also found for the majority of the place fields. An observation is that the prediction quality becomes poor when a barrier directly overlaps with the original place field, leading to a place field that disappears on one side of the barrier (e.g. place cell 3 in Environment 7) or scatters along the end of the environment (e.g. place cell 1 in Environment 7).\\
\newline
Having observed a time series of boundary vector cell firing rates, one can decode the trajectory with fairly high confidence using a trained DNN with the additional pre-processing step, without needing to explicitly record the rate maps and firing rates of the downstream place cells. Of all decoded steps in three cases, it is unclear why the decoding using DNN predictions appear better than using true place cell data, but we suspect this is due to the extra pre-processing step eliminating many of the possible decoded locations. We also note that the decoding quality is restricted by many factors and cannot be fully attributed to the DNN's performance. Firstly, the firing fields of the $10$ simulated place cells do not cover the entire environment, this leads to many locations where no spikes are predicted. Secondly, the MAP estimate is a point estimate and does not convey uncertainty of the posterior distribution. Lastly, the decoding quality is restricted by the spatial resolution as a result of discretisation.

\subsection{Future Works and Improvements}
Our experimental approaches to approximating biological neuronal circuits, albeit simplified and preliminary, have demonstrated promising potentials in the use of modern deep learning techniques in modelling the computations and inferring the cerebral functions encapsulated in these circuits.\\
\newline
Various limiting factors can be identified in the ABNN that deviate from real circuits. Firstly, our input patterns are generated by counting the number of spikes at discrete time bins in simulated spike trains. The mean firing rates of each input neuron is fixed and pre-specified. This simplifying assumption can be extended by adopting dynamic firing rates for each input neuron as (possibly random) functions of an underlying, low-dimensional latent time-series of neural states driven by continuous stimuli \cite{yu2008gaussian}. Secondly, we interpret the output of the ABNN as the (signed) firing rates of each output neuron. In reality, information is encoded within spike timings. To this end, Spiking Neural Networks \cite{maass1997networks} could act as viable alternatives for the ABNNs and the DNNs would predict spike timings and/or intervals upon presenting inputs. Finally, in the plasticity analysis, we model the weight changes in the ABNN as injecting Gaussian noise to weights. While pragmatic, a more realistic scheme would be to present patterns to the ABNN and train it with biologically plausible learning rules, such as the BCM rule \cite{bienenstock1982theory} or Hebb's rule \cite{hebb2005organization}, or in the case of Spiking Neural Networks, with Spike Timing Dependent Plasticity \cite{chakraborty2021characterization}.\\
\newline
Using the trained DNN for the BVC-PC network, we demonstrated that the receptive field of downstream neurons can be predicted with relatively high accuracy. However, it is non-trivial to choose a universal evaluation criterion---in our experiments we used mean squared error loss and Structural Similarity Index. We argued that each criterion has an advantage in its own right, but is nonetheless sensitive to factors resolved by others. For example, MSE loss penalises pixel-wise differences in predicted outputs, but does it take into account translational invariance of place field over the environment. Further, we demonstrated that one can decode the predicted place cell activities to create a cognitive map of the environment. A natural step forward is to test if the trained DNN can fully replace the biological circuit to replicate behaviours in relation to spatial cognition, such as goal-directed reinforcement learning in an environment, similar to experiments conducted by Banino et al. \cite{banino2018vector}.\\
\newline
Modern experimental neuroscience techniques have allowed precision tracing of functional neural pathways, for example, retrograde and anterograde tracing using fluorescence-labelled viruses \cite{saleeba2019student}. Advancements in neuronal population recording equipment such as Neuropixel probes \cite{steinmetz2021neuropixels} have also pushed forward our understandings of neural information processing.  While our experiments have been conducted entirely \textit{in silico}, our ultimate goal is to use a trained DNN fully in place of a biological circuit. Having an accurate DNN as functional connectome of a brain region allows much simpler and non-invasive experimental procedures for neuroscientists, without the need to explicitly record simultaneous neuronal firings. Therefore, future work should also incorporate training the DNN on experimentally recorded firing data, and testing its robustness against stochasticity in neuronal firings, biological change in synaptic strengths, as well as cell deaths. 

\printbibliography

\appendix

\end{document}